\documentclass[prb,reprint,twocolumn,superscriptaddress,noshowpacs,notitlepage,longbibliography,10pt,citeautoscript]{revtex4-2}%

\setcitestyle{super}
\usepackage{graphicx,bm,times}
\graphicspath{ {./figures/main/} }
\usepackage{pgffor}
\usepackage{amsmath}
\usepackage{amsfonts}
\usepackage{amssymb}
\usepackage{mathtools}
\usepackage{color}
\usepackage{hyperref}
\hypersetup{
  colorlinks = true,
  allcolors = {blue}
}
\newcommand{\moire}{moir{\'e} }
\newcommand{\Moire}{Moir{\'e} }
\newcommand{\qms}{\mathbf{q}_\mathrm{m}^{\mathrm{lat}}}
\newcommand{\qmc}{\mathbf{q}_\mathrm{m}^{\mathrm{CDW}}}

\begin{document}

\title{Reconfigurable flat bands from cooperative \moire and charge order}

\author{B.K.~Saika}
\altaffiliation{These authors contributed equally}
\affiliation{SUPA, School of Physics and Astronomy, University of St Andrews, St Andrews KY16 9SS, United Kingdom}

\author{S.~Buchberger}
\altaffiliation{These authors contributed equally}
\affiliation{SUPA, School of Physics and Astronomy, University of St Andrews, St Andrews KY16 9SS, United Kingdom}

\author{S.~Mo}
\affiliation{SUPA, School of Physics and Astronomy, University of St Andrews, St Andrews KY16 9SS, United Kingdom}

\author{A.~Rajan}
\affiliation{SUPA, School of Physics and Astronomy, University of St Andrews, St Andrews KY16 9SS, United Kingdom}

\author{D.~Halliday}
\affiliation{SUPA, School of Physics and Astronomy, University of St Andrews, St Andrews KY16 9SS, United Kingdom}
\affiliation{Diamond Light Source, Harwell Science and Innovation Campus, Didcot, OX11 0DE, United Kingdom}

\author{Y.-C.~Yao}
\affiliation{SUPA, School of Physics and Astronomy, University of St Andrews, St Andrews KY16 9SS, United Kingdom}
\affiliation{Max Planck Institute for Chemical Physics of Solids, Nöthnitzer Strasse 40, Dresden 01187, Germany}

\author{L.C.~Rhodes}
\affiliation{SUPA, School of Physics and Astronomy, University of St Andrews, St Andrews KY16 9SS, United Kingdom}

\author{H.~Decitre}
\affiliation{SUPA, School of Physics and Astronomy, University of St Andrews, St Andrews KY16 9SS, United Kingdom}

\author{T.~Balasubramanian}
\author{C.~Polley}
\affiliation{MAX IV Laboratory, Lund University, Lund, Sweden}

\author{P.~Wahl}
\affiliation{SUPA, School of Physics and Astronomy, University of St Andrews, St Andrews KY16 9SS, United Kingdom}

\author{P.D.C.~King}
\email{pdk6@st-andrews.ac.uk}
\affiliation{SUPA, School of Physics and Astronomy, University of St Andrews, St Andrews KY16 9SS, United Kingdom}

\date{\today}

\begin{abstract}
  The formation of flat electronic bands from long-wavelength superperiodic \moire potentials in van der Waals heterostructures underpins the creation and control of a host of highly-tunable correlated and topological phases.
  While the underlying \moire periodicity is typically a fixed property of the heterostructure, here we show how the development of a charge-density wave (CDW) in one of the constituent materials can create an emergent \moire lattice.
  We demonstrate this experimentally in TiSe$_2$/graphite epitaxial heterostructures, using angle-resolved photoemission and scanning-tunnelling microscopy and spectroscopy to directly image the resulting long-wavelength \moire potential and concomitant flat-band formation.
  We show how the intrinsically low-energy, deformable landscape of the CDW imparts significant tunability, stabilising quasi-one-dimensional \moire domains from symmetry breaking within the CDW, and allowing the complete suppression of flat-band formation by carrier doping across the CDW phase transition.
  Our findings thus open a new avenue for engineering \moire matter by exploiting the rich many-body states of the parent compounds of 2D heterostructures.

\end{abstract}

\maketitle

Stacking two-dimensional materials to form van der Waals (vdW) heterostructures has created an exciting platform for realising quantum metamaterials with precisely engineered electronic  states~\cite{cao_correlated_2018,cao_unconventional_2018,andrei_marvels_2021,kennes_moire_2021,mak_semiconductor_2022,nuckolls_microscopic_2024}.
Freed from the constraints of lattice-matching heterointerfaces in traditional covalent materials, vdW heterostructures can be created with a near arbitrary twist or lattice mismatch between the constituent layers~\cite{geim_van_2013}.
This naturally induces a new periodic, so-called moir{\'e}, potential in the system~\cite{bistritzer_moire_2011,wu_topological_2017}, whose wavelength can be tuned by the twist angle or lattice constants of the constituent materials.
Extremely flat electronic bands can result~\cite{bistritzer_moire_2011,cao_correlated_2018,utama_visualization_2021,lisi_observation_2021,angeli__2021,pei_observation_2022,gatti_flat_2023}, leading to a host of strongly-correlated phases including Chern, Mott, and other correlated insulators~\cite{cao_correlated_2018,chen_evidence_2019, regan_mott_2020,tang_simulation_2020,li_quantum_2021}, unconventional magnets~\cite{sharpe_emergent_2019,chen_tunable_2020, serlin_intrinsic_2020,tschirhart_imaging_2021}, superconductors~\cite{cao_unconventional_2018,yankowitz_tuning_2019,lu_superconductors_2019}, and other broken-symmetry states~\cite{saito_hofstadter_2021,liu_visualizing_2022,yu_correlated_2022}.
\Moire heterostructures, therefore, in principle provide model platforms for realising truly designer quantum materials.
However, their dynamical control remains a significant challenge, where modifying the underlying \moire periodicity typically necessitates fabrication of new samples with different twist angles or lattice mismatch.

\begin{figure*}
  \centering
  \includegraphics[width=0.95\textwidth]{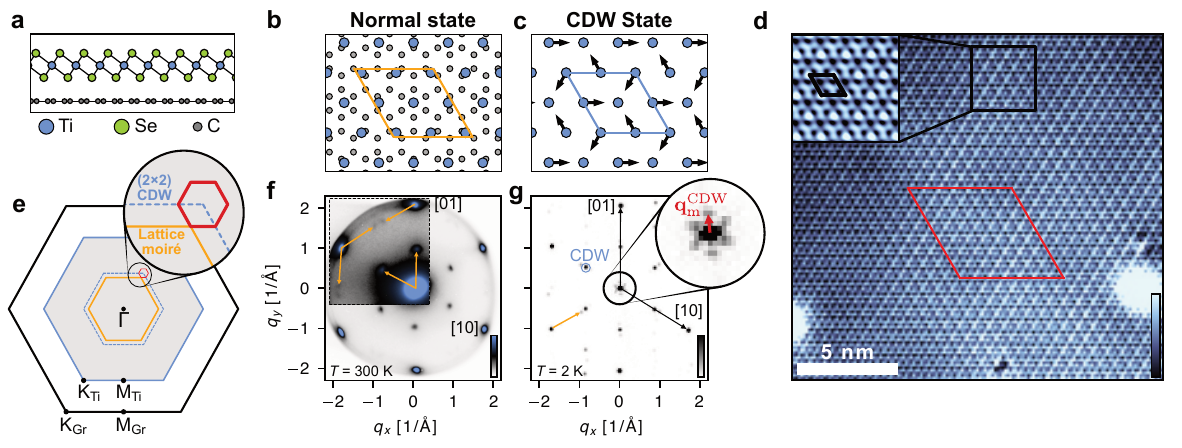}
  \caption{{\bf Hybrid CDW-\moire lattice formation in monolayer (ML)-TiSe$_2$/graphite van der Waals heterostructures.}
    (a) Schematic side view of ML-TiSe$_2$ grown epitaxially on graphite.
    (b) Top view of the resulting \moire superlattice formed from the lattice mismatch between the top-most graphite and the $(1\times1)$ undistorted (normal-state) TiSe$_2$ (yellow: \moire unit cell).
    (c) Ti atom displacements (arrows) in the CDW state, showing the resulting $(2\times2)$ unit cell.
    (d) STM topography in the CDW state ($T=1.8$~K, tunnelling setpoint $V_{s} = 220$ mV, $I_{s}=100$ pA).
    The inset shows a high-resolution magnified view highlighting the CDW $2 \times 2$ periodicity (tunnelling setpoint $V_{s} = 60$ mV, $I_{s}=420$ pA).
    (e) Brillouin zones corresponding to the graphite (black), $(1\times1)$ TiSe$_2$ (blue), normal-state lattice \moire (yellow) and CDW (dashed blue) unit cells.
    A putative secondary \moire zone is shown in red, from the interplay of the latter two.
    (f) Low-energy electron diffraction pattern measured in the normal state of TiSe$_2$ ($T=300$~K) using a low-energy electron microscope (see Methods).
    The inset shows the measured pattern with enhanced contrast, highlighting Bragg peaks from the substrate-imposed \moire lattice (yellow arrows).
    (g) Fourier transform of the STM topography in (d) showing the primary TiSe$_2$ Bragg peaks together with an array of charge-order ($\mathbf{q}_{\mathrm{CDW}}$) and substrate-induced \moire ($\qms$) peaks, as well as the co-operative effect of both ($\qmc$, see inset).
    The long-wavelength real-space modulation corresponding to $\qmc$ is also visible in our topographic measurements (d, red unit cell).
  }
  \label{fig:overview}
\end{figure*}

Here, we demonstrate a new approach: stabilising a tunable \moire periodicity from the cooperative effect of a fixed lattice-driven \moire potential and the formation of a charge-density wave (CDW) -- a broken symmetry state where the electron fluid develops its own periodic modulation -- in one of the constituent layers of the heterostructure.
We show this possibility in epitaxial van der Waals heterostructures formed by the growth (see Methods) of monolayer (ML) TiSe$_2$ on a graphite substrate (Fig.~\ref{fig:overview}(a)).
The lattice mismatch between the graphite and the TiSe$_2$ (Fig.~\ref{fig:overview}(b)) creates a fixed \moire potential with wavevector
\begin{equation}
  |\qms|=\frac{4\pi}{\sqrt{3}}\left(\frac{a_{\mathrm{TiSe_2}}-a_\mathrm{Gr}}{a_{\mathrm{TiSe_2}}\cdot a_\mathrm{Gr}}\right),
  \label{eq:moire_vector}
\end{equation}
where $a_{\mathrm{TiSe_2}}$ and $a_{\mathrm{Gr}}$ are the in-plane lattice constants of the TiSe$_2$ and graphite (Gr) layers, respectively. A substantial lattice mismatch of $(a_{\mathrm{TiSe_2}}-a_\mathrm{Gr})/{a_{\mathrm{TiSe_2}}}\approx\!\!30$~\% means that the associated \moire wavevector is large, $|\qms|=0.90~\text{\AA}^{-1}$, remaining on the same order as the one associated with the atomic lattice.
ML-TiSe$_2$, however, also hosts an instability to a $(2\times2)$ CDW state (Fig.~\ref{fig:overview}(c,d))~\cite{di_salvo_electronic_1976,peng_molecular_2015, chen_charge_2015,watson_strong-coupling_2020,buchberger_persistence_2025}.
The resulting reduced Brillouin zone lies close to that of the \moire zone formed by the substrate mismatch with the $(1\times1)$ unreconstructed unit cell of TiSe$_2$ (Fig.~\ref{fig:overview}(e)), opening the possibility to stabilise a much longer-wavelength \moire potential from the interplay between these two orders.
We demonstrate this here, uncovering both the formation of flat electronic bands from the corresponding emergent \moire periodicity and revealing their intrinsic tunability arising from the shallow energy landscape of the CDW order.

\subsection*{Hybrid \moire lattice}

Figure~\ref{fig:overview}(f) shows room-temperature electron diffraction measurements from our ML-TiSe$_2$/Gr heterostructures, measured in the normal state of TiSe$_2$.
As well as the Bragg spots of the TiSe$_2$ lattice (e.g., labelled $[10]$ and $[01]$), additional weaker diffraction features are observed, separated from the primary Bragg spots by $|\mathbf{q}|=0.93\pm{0.02}$~\AA$^{-1}$ (yellow arrows).
This is very close to $|\qms|$ expected from the lattice mismatch of the TiSe$_2$ and graphite layers (Eq.~\ref{eq:moire_vector}, see also Supplementary Information, Section~I), confirming that the finite lattice mismatch between the substrate and epilayer leads to a well-defined, if rather short-wavelength, \moire heterointerface in the normal state of TiSe$_2$.

In Fig.~\ref{fig:overview}(d), we show scanning tunnelling microscopy (STM) measurements of the heterostructure taken at $1.8$~K.
A strong $(2\times2)$ modulation of the atomic contrast (see inset) is consistent with the known~\cite{peng_molecular_2015, chen_charge_2015,watson_strong-coupling_2020,buchberger_persistence_2025} CDW instability in ML-TiSe$_2$.
This gives rise to well-defined half-order Bragg peaks in Fourier transform analysis of our measured STM data (Fig.~\ref{fig:overview}(g)).
In addition, however, we find a rich array of weaker satellite peaks.
Most notably, these include a hexagon of Bragg peaks at small wavevectors, which are well described by $\qmc=\mathbf{q}_{\mathrm{CDW}}-\qms$, with $|\qmc|=0.14$~\AA$^{-1}$.
This additional periodicity is also visible as a periodic modulation of the intensity in our measured topography shown in Fig.~\ref{fig:overview}(d), with a relatively long wavelength of $>5$~nm.
This directly indicates the formation of a hybrid \moire lattice here, arising from the interplay of the normal-state TiSe$_2$/Gr lattice mismatch and the charge order of the TiSe$_2$ layer (Fig.~\ref{fig:overview}(e), also supported by model calculations following the \moire lattice construction of Zeller and G{\"u}nther~\cite{zeller_what_2014} (Supplementary Information, Section I)).

\subsection*{Spectroscopic observations of flat-band formation}
\begin{figure*}
  \centering
  \includegraphics[width=0.95\textwidth]{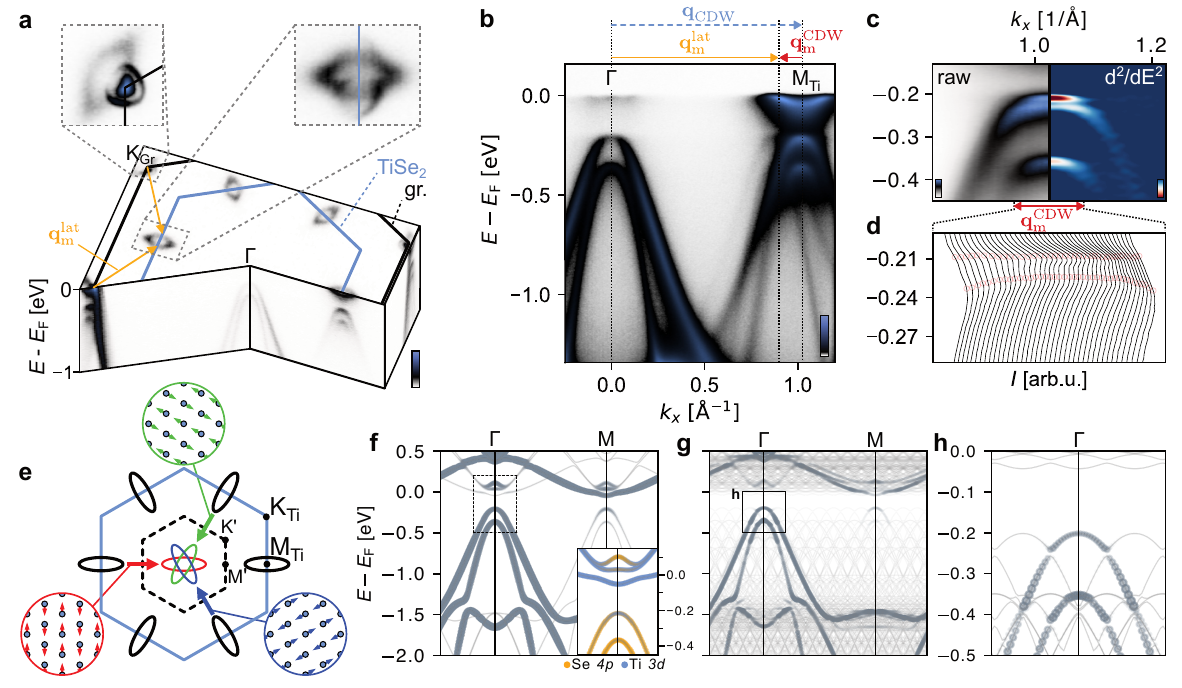}
  \caption{{\bf Formation of flat electronic bands.}
    (a) Overview of the electronic structure of ML-TiSe$_2$/Gr, measured by ARPES ($h\nu=150$~eV, linear horizontal (LH)-polarisation, $T=18$~K).
    The insets show magnified views of the Fermi surface measured in the vicinity of the graphite K-point ($h\nu=80$~eV, LH-pol) and M-point of the TiSe$_2$ zone ($h\nu=37$~eV, LH-pol), respectively.
    (b) Dispersion measured along $\Gamma$-M ($h\nu=30$~eV, LH-pol.), showing the primary conduction and valence bands, their CDW replica, and additional \moire replica (see coloured arrows).
    (c) Magnified view ($h\nu=20$~eV, LH-pol., left) and corresponding second-derivative analysis (right) close to the top of the valence band backfolded by the CDW.
    (d) Selected EDCs extracted from the measured dispersion close to M$_\mathrm{Ti}$, and corresponding fitted peak positions (open circles, see Supplementary Fig.~S2), showing the formation of flat electronic bands.
    (e) Schematic illustration of backfolding of three elliptical conduction band pockets by the $3\mathbf{q}$ CDW instability in TiSe$_2$.
    (f) Tight-binding modelling of TiSe$_2$ band structure in the CDW state.
    The inset shows the atomic character of the resulting hybridised bands.
    (g) Calculations additionally incorporating a periodic potential defined by the hybrid CDW-\moire wavevector $\qmc$, projected onto the states of the original cell and
    (h) zoomed view of valence dispersion with the flat-band formation at the valence band top.
  }
  \label{fig:flat_bands}
\end{figure*}

Fig.~\ref{fig:flat_bands}(a) shows an overview of the electronic structure of the heterostructure as measured by angle-resolved photoemission (ARPES).
The $\pi$ states of graphite are visible at the zone-corner K-point of the graphite Brillouin zone, with a hole-doped split-off copy evident due to charge transfer from the graphite to the TiSe$_2$ layer (see left inset).~\cite{mo2026resonant,buchberger_persistence_2025}
Replicas of these graphite pockets are visible close to, but slightly offset from, M$_\mathrm{Ti}$ in the vicinity of the TiSe$_2$ conduction bands (right inset in Fig.~\ref{fig:flat_bands}(a)).
We attribute these as \moire replica, shifted from K$_\mathrm{Gr}$ by the wavevector $\qms$ of the \moire potential created by the substrate-epilayer lattice mismatch.~\cite{mo2026resonant}
Similar \moire replica can be observed for the TiSe$_2$ states (Fig.~\ref{fig:flat_bands}(b)).
Together with the Se 4$p$-derived valence bands ``backfolded'' from $\Gamma$ to M$_\mathrm{Ti}$ by the $(2\times2)$ CDW~\cite{rossnagel_charge-density-wave_2002,cercellier_evidence_2007,watson_orbital-and_2019,chen_charge_2015,watson_strong-coupling_2020,buchberger_persistence_2025},
we also observe weak but clear signatures of additional offset copies of these bands shifted away from M$_\mathrm{Ti}$ by the hybrid \moire wavevector $\qmc$.

\begin{figure*}
  \centering
  \includegraphics[width=0.95\textwidth]{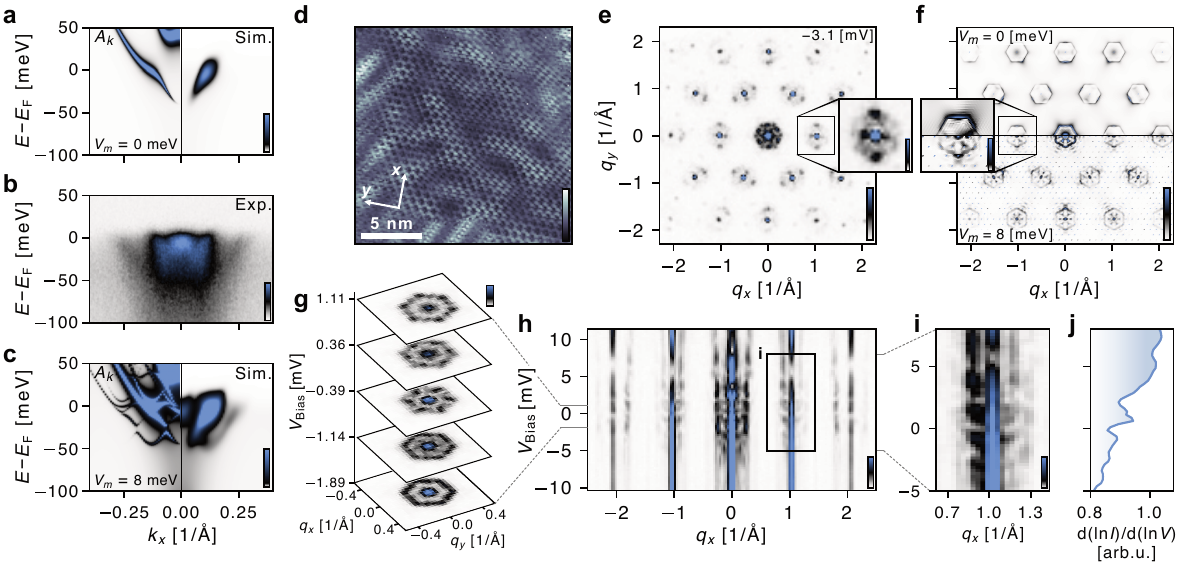}
  \caption{{\bf Complex Fermiology from multi-band \moire coupling.}
    (a) Unfolded spectral function calculations (left) and simulated ARPES spectra (right, $T=18~\mathrm{K}$; $7~\mathrm{meV}$ broadening) in the vicinity of $E_\mathrm{F}$ from our CDW tight-binding model Hamiltonian.
    (b) Corresponding ARPES measurements of the conduction band at $\Gamma$ ($h\nu = 25$~eV, LH-pol.), showing significant additional complexity.
    (c) This is captured in calculations which additionally include a hybrid CDW-\moire potential (as (a) but with $V_{m}=8~\mathrm{meV}$).
    (d) Low-temperature spatially-resolved $\mathrm{d}I/\mathrm{d}V$ conductance maps of ML-TiSe$_2$/Gr at $-3.14$~mV ($V_{s} = 11.4$ mV, $I_{s}=800$ pA, $V_{\mathrm{lock-in}}=0.25~\mathrm{meV}$).
    (e) Corresponding symmetrised Fourier transform (see Supplementary Fig.~S7) showing strong QPI features around CDW and the Bragg peaks (see inset).
    (f) Simulated QPI at $E-E_\mathrm{F}=-3.0~\mathrm{meV}$ obtained from calculations without (top, $V_{m}=0$~meV) and with (bottom, $V_{m}=8$~meV) the hybrid CDW-\moire potential. Inset shows corresponding features around CDW peak.
    (g) QPI contours taken at different energies, showing rapid changes across a narrow window of 3 meV.
    (h,i) Corresponding QPI dispersions along $\Gamma-\mathrm{M_{Ti}}$ (h) and magnified view around the CDW peak (i), showing the formation of ultra-flat bands.
    (j) These are visible as sharp peaks in the corresponding average $\mathrm{d}(\ln{I})/\mathrm{d(\ln{V})}$ spectrum.
  }
  \label{fig:3q_qpi}
\end{figure*}

In the vicinity of where these \moire and CDW replicas intersect, we find clear spectroscopic signatures of extreme band flattening (Fig.~\ref{fig:flat_bands}(c,d)).
A pair of flat states are formed at the valence band top.
From fitting to extracted energy distribution curves (EDCs, Fig.~\ref{fig:flat_bands}(d), see also Supplementary Information, Section II) we estimate that the uppermost of these has an ultra-narrow bandwidth of only $(2^{+3}_{-1})$~meV.
This is far narrower than any of the electronic states expected for a hypothetical isolated TiSe$_2$ layer.\cite{watson_strong-coupling_2020}
We attribute this to the presence of the hybrid lattice and charge-order \moire potential.

To validate this, we calculate the electronic structure of the hybrid \moire system starting from a tight-binding description of the CDW state (see Methods)~\cite{kaneko_exciton-phonon_2018}.
Here, we consider the CDW as a $3\mathbf{q}$ instability, resulting from the superposition of three sets of unidirectional atomic displacements as shown in Fig.~\ref{fig:flat_bands}(e).
This backfolds the conduction bands from the three M$_\mathrm{Ti}$-points to $\Gamma$, where they hybridise with the valence bands in an orbitally-selective manner (Fig.~\ref{fig:flat_bands}(f))~\cite{watson_orbital-and_2019,bianco_electronic_2015,kaneko_exciton-phonon_2018,monney_spontaneous_2009,watson_strong-coupling_2020,antonelli_orbital-selective_2022}.
We additionally incorporate the hybrid CDW-\moire via inclusion, on all orbitals, of the first-order harmonic of an on-site scalar potential (see Methods).
This leads to a significant reconstruction of the electronic structure (Fig.~\ref{fig:flat_bands}(g)),
including the formation of a large number of replica bands and the opening of hybridisation gaps where these cross.

This hybridisation yields a narrow state at the top of the valence band (Fig.~\ref{fig:flat_bands}(h)), qualitatively consistent with our experimental measurements.
We note that only a single mini-band is predicted close to the valence band top, in contrast to the pair of flat features observed experimentally.
We will show below that this reflects the formation of domains in the CDW state (see also Supplementary Information, Section III).
The model also overestimates the bandwidth of the flat band(s) that are formed.
This may result from an overestimation of the valence band curvature in the parent model due to an underestimation of the CDW-induced hybridisation (see Supplementary Information, Section IV, Fig.~S6) or the neglect of higher-order hopping terms.
Furthermore, the model employs an orbital- and atom-independent magnitude of the \moire potential which is likely an oversimplification, in particular as the CDW-induced band hybridisations lead to orbital compositions which vary strongly on momentum scales comparable to the size of the \moire zone (Fig.~\ref{fig:flat_bands}(f, inset)).

\begin{figure*}
  \centering
  \includegraphics[width=\textwidth]{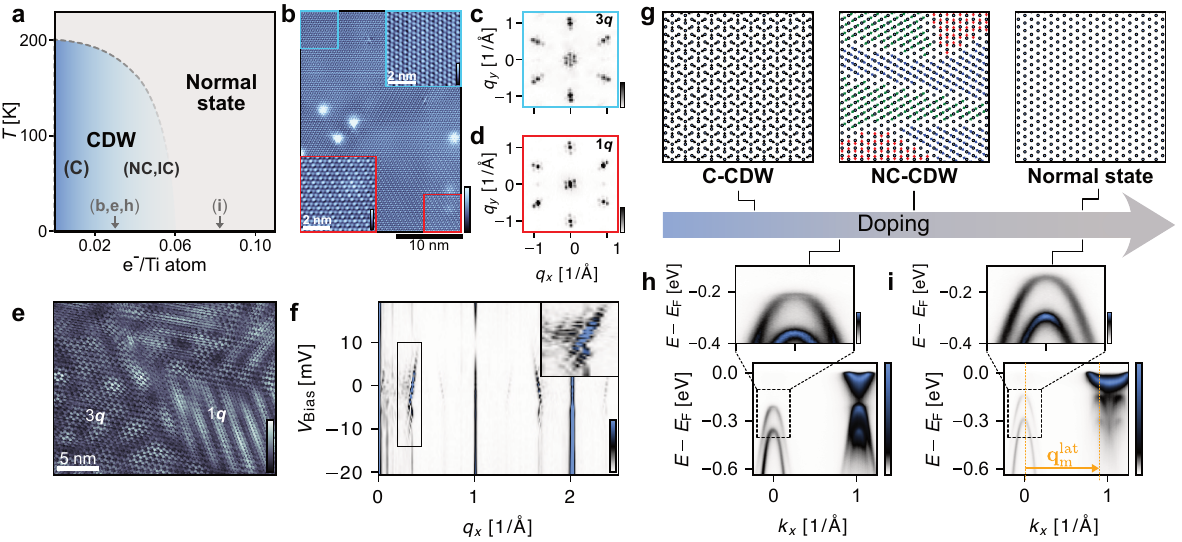}
  \caption{{\bf \Moire tunability via low-energy control of the CDW.}
    (a) Schematic doping-dependent CDW phase diagram of ML-TiSe$_2$, showing the transition from commensurate (C-) to nearly commensurate (NC-) and incommensurate (IC-) order with increasing electron doping, before the CDW becomes completely suppressed.
    (b) NC-CDW domains as visualized in STM topography (tunnelling setpoint $V_{s} = 100$ mV, $I_{s}=100$ pA) showing regions of dominant single-$\mathbf{q}$ ($1\mathbf{q}$) and triple-$\bf \mathbf{q}$ ($3\mathbf{q}$) CDW order.
    (c, d) FFT of regions with dominant (c) $1\mathbf{q}$ and (d) $3\mathbf{q}$ order.
    The latter shows anisotropic hybrid CDW-\moire peaks.
    (e) STM $\mathrm{d}I/\mathrm{d}V$ conductance map at $-0.64$~meV ($V_{s} = 19.4$ mV, $I_{s}=650$ pA, $V_{\mathrm{lock-in}}=1~\mathrm{mV}$).
    (f) High-resolution QPI dispersions along $\Gamma-\mathrm{M_{Ti}}$ extracted from a region with dominant $1\mathbf{q}$ phase ($V_{s} = 19.4$ mV, $I_{s}=600$ pA, $V_{\mathrm{lock-in}}=0.5~\mathrm{mV}$, see Supplementary Fig.~{S4}).
    (g) Schematic illustration of the doping evolution of the lattice configuration from C-CDW to NC-CDW to the normal state.
    (h, i) Overview of the corresponding ARPES measurements along $\Gamma$-M ($h\nu=39$~eV, LV-pol.) and (inset) of the valence band at $\Gamma$ ($h\nu=17.5$~eV, LH-pol.).
    These are shown from an as-grown sample ((h), as in (e)), thus averaging over $3\mathbf{q}$ and $1\mathbf{q}$ domains and (i) following deposition of dilute (sub-ML) coverage of Rb atoms on the surface.
    This causes an electron doping and the suppression of the CDW and the flat-band formation.
  }
  \label{fig:CDW_domains}
\end{figure*}

Indeed, such complexity of the starting electronic structure is a notable feature of the present system as compared to established \moire materials.~\cite{tang_simulation_2020,angeli__2021,claassen_ultra-strong_2022}
This is of particular interest for the conduction bands, where the CDW-induced orbital mixing leads to multiple highly non-parabolic states forming in the vicinity of the band bottom (Fig.~\ref{fig:flat_bands}(f), inset).
In the absence of the \moire potential, only the bottom-most split-off branch of these would be expected to cross $E_\mathrm{F}$ at the experimental doping level (Fig.~\ref{fig:3q_qpi}(a)).
In contrast, our ARPES measurements (Fig.~\ref{fig:3q_qpi}(b)) indicate a complex Fermiology which is well-described by our calculations including the hybrid CDW-\moire potential (Fig.~\ref{fig:3q_qpi}(c)).

Our experiment and calculations thus indicate a rich hierarchy of intersecting \moire mini-bands forming in the vicinity of the Fermi level.
To better resolve these, we turn to high-resolution scanning-tunnelling spectroscopy (STS, Fig.~\ref{fig:3q_qpi}(d-j), see Methods).
Here, strong signatures of quasiparticle interference (QPI) are observed in real-space (Fig.~\ref{fig:3q_qpi}(d)), giving rise to clear but complex scattering patterns in Fourier space around both the structural and CDW-derived Bragg peaks (Fig.~\ref{fig:3q_qpi}(e)).
The observed QPI patterns exhibit strongly modulated intensities with sharp vertex-like features, in contrast to the dominant hexagonal patterns expected from QPI simulations of the unreconstructed band structure (Fig.~\ref{fig:3q_qpi}(f,top)).
Including the CDW-\moire potential in our simulations (Fig.~\ref{fig:3q_qpi}(c, f-bottom)) naturally reproduces the sharp features observed experimentally as a result of additional intra- and inter-band scattering channels opened through the underlying \moire band reconstruction.
Consistent with this, we find that the experimental QPI patterns are extraordinarily sensitive to small changes in bias voltage (Fig.~\ref{fig:3q_qpi}(g), see also Supplementary Video~S1).
This points to the formation of flat bands and the opening of hybridisation gaps that modulate the available quasiparticle scattering channels on small energy scales.
This is also evident in our QPI dispersions extracted along $\Gamma-\mathrm{M_{Ti}}$ (Fig.~\ref{fig:3q_qpi}(h,i)).
A series of discontinuities (Fig.~\ref{fig:3q_qpi}(i)) point to ultra-flat QPI dispersions, indicating the formation of flat bands with widths on the order of only 2-3~meV, leading to a sequence of sharp peaks in the density of states, evident in our normalised conductance measurements ($\mathrm{d}(\ln{I})/\mathrm{d}(\ln{V})$, Fig.~\ref{fig:3q_qpi}(j)).

\subsection*{Low-energy CDW control of the \moire lattice}
Our results above therefore establish a new modality for creating \moire potentials -- and associated ultra-flat electronic bands -- via a co-operative effect of lattice mismatch and charge order.
Excitingly, this opens routes to their low-energy control.
Fig.~\ref{fig:CDW_domains}(a) shows the CDW phase diagram of TiSe$_2$, established from previous studies of bulk and monolayer samples~\cite{joe2014emergence,kogar2017observation,yan2017influence,watson_strong-coupling_2020,munoz2025doping,wan_near-commensuration_2026}.
As the electron doping is increased, an incommensuration of the $3\mathbf{q}$
CDW occurs.~\cite{joe2014emergence, li_controlling_2015, kogar2017observation}
Recent studies of ML-TiSe$_2$ have shown how this proceeds via a nearly-commensurate (NC) phase, accompanied by the formation of nematic domains~\cite{munoz2025doping,wan_near-commensuration_2026}.
Interestingly, our STM measurements (Fig.~\ref{fig:CDW_domains}(b)) reveal domains of a stripe-type order favouring a dominant single $\mathbf{q}$ ($1\mathbf{q}$) CDW vector in addition to patches of the $3\mathbf{q}$ order discussed above.
This places our samples in the vicinity of the NC-CDW phase~\cite{wan_near-commensuration_2026}, likely as a result of the inherent charge transfer from the substrate to the TiSe$_2$ layer.

The Fourier transforms shown in Fig.~\ref{fig:CDW_domains}(c, d) reveal how the presence of a quasi-1D CDW order in our samples directly translates into the formation of an anisotropic hybrid CDW-\moire potential (see anisotropy of low-$\mathbf{q}$ Bragg peaks in Fig.~\ref{fig:CDW_domains}(d) vs.\ (c)).
The reconstructed electronic structure responds sensitively, as evident from the emergence of strongly unidirectional QPI in the $1\mathbf{q}$ CDW regions (Fig.~\ref{fig:CDW_domains}(e)).
A high resolution $(\mathrm{d}I/\mathrm{d}V)$ QPI dispersion from such a stripe-type domain is shown in Fig.~\ref{fig:CDW_domains}(f).
The QPI of $1\mathbf{q}$ CDW regions still reveal discontinuities in the conduction band originating from replica formation and associated band hybridisation introduced by the \moire potential (see inset).
However, unlike regions with $3\mathbf{q}$ CDW order, the extreme band flattening near the Fermi level is not present.
This points to a profound influence of \moire dimensionality on the hybridisation schemes which result.

Notably, this should be tunable via charge carrier doping (Fig.~\ref{fig:CDW_domains}(g)), for example using an electrostatic gate~\cite{li_controlling_2015} to deplete carriers from the TiSe$_2$ layer to stabilise a robust $3\mathbf{q}$ phase, or by electron doping to drive the system towards the IC phase.
At the native substrate-induced charge-transfer doping here, the system is in the vicinity of the NC phase and spatial domains of both $3\mathbf{q}$ and $1\mathbf{q}$ order co-exist.
As well as hosting distinct \moire hybridisation of the conduction bands, tunnelling spectra obtained from the $1\mathbf{q}$ and $3\mathbf{q}$ regions exhibit an approximately 25~meV shift of a peak in the local density of states at the valence band top (Supplementary Fig.~S3).
This provides a natural explanation for the pair of flat-topped valence bands observed in our ARPES measurements, discussed above and visible in the inset of Fig.~\ref{fig:CDW_domains}(h), which will integrate over both types of domains.

With further increase in electron doping, the CDW is expected to be completely suppressed (Fig.~\ref{fig:CDW_domains}(a, h)).~\cite{morosan_superconductivity_2006,li_controlling_2015,watson_strong-coupling_2020}
We achieve this here by the deposition of dilute concentrations of Rb atoms at the surface (see Methods).~\cite{bostwick_quasiparticle_2007,kim_observation_2015,riley_negative_2015}
From the increased size of the electron pocket following Rb deposition ({\it cf.} Fig.~\ref{fig:CDW_domains} (h) and (i)), we estimate that this yields an electron density in the TiSe$_2$ layer of $\sim0.08$ $e^{-}$/Ti.
An upward shift of the valence band maximum and suppression of the backfolded valence band spectral weight at M$_\mathrm{Ti}$ indicates a strong suppression of the CDW.
Well-defined replicas of the valence bands are still observed close to, but slightly offset from, the M$_{\mathrm{Ti}}$ point, indicating that the large-$q$ \moire potential arising from the direct inter-layer lattice mismatch remains.
However, the small-$\mathbf{q}$ hybrid CDW-\moire potential must be suppressed as the CDW is destroyed.
Importantly, our corresponding measurements of the valence-band dispersion (Fig.~\ref{fig:CDW_domains}(i, inset)) now show a complete suppression of the flat-band formation at the band top, with conventional near-parabolic Se-derived bands restored.

\subsection*{Discussion and Outlook}
Our measurements therefore confirm the inherently co-operative nature of the long-wavelength \moire potential here, arising from the interference between the periodic potentials from the charge and original lattice-induced \moire order.
The fact that the formation of the long-wavelength \moire potential itself depends on a collective quantum phase opens new prospects for its control.
To this end, we have shown here how an emergent nematicity of the CDW in the vicinity of the commensurate-to-incommensurate transition dramatically reshapes the \moire potential and the band hybridisation which results.
Depleting carriers to stabilise the commensurate phase will lead to the most robust flat-band regime, where small further changes in carrier density can be used to tune the chemical potential within the flat bands, potentially triggering new collective state formation.
Meanwhile, deliberate electron doping can be used to controllably deactivate the long-wavelength \moire potential, enabling dynamical reconfiguration of the \moire landscape.
In the bulk, as the CDW order is driven into the incommensurate phase, superconductivity is known to emerge.~\cite{morosan_superconductivity_2006}
It will be of interest to investigate the interplay of such a superconducting state with the \moire-reconstructed electronic structure in this crossover regime.

Our materials platform investigated here also opens routes for chemical control of the \moire order.
The lattice constant of the TiSe$_2$ layer can be increased by substituting Te for Se, with a $(2\times2)$ charge order remaining stable over the entire compositional range~\cite{antonelli_controlling_2023}.
This opens the tantalising prospect to continuously tune the wavelengths of the substrate-induced \moire and the charge order, which should become coincident at a critical composition of TiSe$_{0.8}$Te$_{1.2}$.
The resulting wavelength of the hybrid CDW-lattice \moire should therefore also be tunable, diverging close to the critical composition.
This provides a new platform for highly controlled studies of the impact of long-wavelength \moire order on the electronic and collective states.
More generally, our study outlines a broader potential for utilising collective states to dynamically tune the potential landscape imposed by the fixed lattice geometry in 2D van der Waals heterostructures, opening new prospects for designer \moire matter.

\

{\small
  \section*{Acknowledgements}
  We thank Brice Sarpi for significant assistance with the $\mu$-LEED experiments at Diamond Light Source.
  We also thank Bernd Braunecker, Carolina A.\ Marques, Kai Rossnagel, and David T.\ S.\ Perkins for useful discussions, and Martin McLaren for technical support.
  We gratefully acknowledge support from the UK Engineering and Physical Sciences Research Council (Grant Nos.~EP/X015556/1, EP/M023958/1, and EP/R025169/1).
  B.K.S.\ also acknowledges support from the Horizon Europe research and innovation funding programme under the Marie Sk{\l}odowska-Curie Actions grant ID 101202935.
  We thank MAX IV Laboratory for time on the Bloch beamline under Proposal No.~20241124. Research conducted at MAX IV, a Swedish national user facility, is supported by the Swedish Research council under contract 2018-07152, the Swedish Governmental Agency for Innovation Systems under contract 2018-04969, and Formas under contract 2019-02496.
  We thank Diamond Light Source for access to the offline I06 PEEM instrument (Proposal MM34231), which contributed to the results presented here.
  The laserPEEM/LEEM offline facility has been partly funded by the EPSRC grant (EP/V029258/1).
  For the purpose of open access, the authors have applied a Creative Commons Attribution (CC BY) licence to any Author Accepted Manuscript version arising.
  The research data supporting this publication can be accessed at [[DOI TO BE INSERTED]].

  \section*{Methods}
  \noindent{\bf Molecular beam epitaxy:} Monolayer TiSe\textsubscript{2} films were grown by molecular beam epitaxy (MBE) in a DCA R450 MBE growth reactor on natural graphite crystals purchased from \textit{NGS Natugraphit} and on 1~mm thick highly oriented pyrolytic graphite (HOPG) substrates purchased from \textit{MaTeck}.
  For easier handling, the natural graphite crystals were glued onto 1\,mm thin tantalum foil chips using high-temperature carbon paste from Agar Scientific.
  The glue was outgassed at $>\!600^\circ\text{C}$ under $\sim\!1\times10^{-7}\,\text{mbar}$ vacuum in a separate annealing chamber.
  The substrates were cleaved in air using adhesive tape to expose a fresh surface  and loaded immediately into the load lock of the growth vacuum system, where they were degassed at $200^\circ\text{C}$ for 10\,h.
  Directly before the film growth, the substrates were annealed at $750^\circ\text{C}$ for $30\,$minutes.

  All samples were grown using a nucleation-assisted MBE growth mode as introduced in Refs.~\cite{Rajan2024,rajan_substrate_2025}
  The films for photoemission measurements were grown on natural graphite, using a growth temperature of $550^\circ\text{C}$.
  3N5 pure Ti was evaporated from a high-temperature effusion cell at a cell temperature of $1455^\circ\text{C}$.
  A valved cracker cell was used for the evaporation of 5N pure Se.
  The base of the cell was held at a temperature of $163^\circ\text{C}$ and the cracking zone at $500^\circ\text{C}$ to produce a supply of Se monomers and dimers with a beam-equivalent pressure of $2\times10^{-7}$\,mbar.
  The low Ti supply used for the growth hinders the accurate determination of the Ti beam-equivalent pressure, but it is estimated to be at least two orders of magnitude lower than the Se supply.
  For an enhanced nucleation rate,~\cite{Rajan2024} 5N pure Ge was co-evaporated from a FOCUS EFM electron beam evaporator with a flux of 1.0\,nA as measured by the integrated flux monitor.
  With this method, a near-complete monolayer film was deposited within a growth time of 1\,h.
  To end the growth, the supply of Ti and Ge was cut off and the sample temperature was ramped down at a rate of 20\,K/min under Se flux to avoid excessive Se vacancy formation.
  The Se supply was cut off at $260^\circ\text{C}$.
  To protect the samples during the transport to the synchrotron, they were capped with a $\sim 5\,\text{nm}$ thick amorphous Se layer, deposited at room temperature.
  The samples were then taken out of the vacuum system and transferred to the beamline.
  At the Max-IV synchrotron, the samples were decapped by annealing them at $260^\circ\text{C}$ under $\sim10^{-10}\,\text{mbar}$ vacuum for $1\,\text{h}$ in a preparation chamber connected to the measurement chamber.

  HOPG substrates were used for the samples for scanning tunnelling microscopy (STM) and micro-focus low-energy electron diffraction ($\mu$-LEED).
  For the STM data shown here, the enhanced nucleation was achieved by pre-sputtering the substrate with He ions~\cite{rajan_substrate_2025}
  (partial pressure: $5.6\times10^{-7}$~mbar), using an accelerating voltage of 200\,V and for a duration of 30\,s.
  In turn, no Ge was co-evaporated during the growth.
  A slightly shorter growth time of 55\,min was used, while all remaining growth parameters were as described above for the photoemission samples.
  We also fabricated samples using the Ge-assisted growth method without additional pre-sputtering, achieving very similar results.
  For the samples for $\mu$-LEED, we used a growth temperature of $600^\circ\text{C}$, Ti cell temperature of $1460^\circ\text{C}$, Se beam equivalent pressure of $2\times10^{-7}$\,mbar, a Ge flux of 1.0\,nA for assisted nucleation and a growth time of 57\,minutes.
  The samples were transferred from the growth chamber to the STM and $\mu$-LEED measurement chambers under ultra-high vacuum, using vacuum suitcases.

  We note that we observe a small rotation of the \moire lattice relative to the TiSe$_{2}$ peaks in our experimental measurements (Fig.~\ref{fig:overview}(g)).
  The \moire lattice orientation is highly sensitive to the relative angle between the film and substrate~\cite{zeller_what_2014}.
  In the measurements shown in Fig.~\ref{fig:overview}(d, g), we estimate the twist angle between TiSe$_{2}$ and graphite to be on the order of $0.2^\circ$, although we find some spatial and sample-to-sample variation in the precise angle.
  This has, however, only a small impact on $|\qmc|$, and so we neglect the finite rotation in our discussions.

  \

  \noindent{\bf Angle-resolved photoemission:} ARPES experiments were performed at the Bloch endstation (A-branch) of the MAX IV synchrotron, Sweden, using a Scienta DA30 electron analyser.
  The utilised photon energies and light polarisations are noted in the figure captions, and the sample was kept at the base temperature of $T=18$~K throughout.
  The light spot size on the sample was $\approx\!10$~$\mu$m$~\times~15$~$\mu$m.
  Surface dosing was performed in the analysis chamber using a SAES Rb alkali dispenser held at a current of 4.5\,A, with the sample held at 18\,K.
  The current on the source was ramped up before the sample was rotated to face the doser, with a deposition time of 5\,s as measured from the point at which the sample was directly facing the source.
  The carrier densities of pristine and dosed samples were determined from a Luttinger analysis of their measured Fermi surfaces.
  The carrier densities of pristine samples were found to be in the range 0.03-0.04\,e\textsuperscript{-} per Ti atom.
  After a single 5\,s dose of Rb deposition, these increased to 0.08-0.09\,e\textsuperscript{-} per Ti atom.
  In addition, we performed a second dosing of 30\,s for the data shown in Supplementary Fig.~S4, from which we estimate a carrier density of 0.14\,e\textsuperscript{-} per Ti atom.
  Additional ARPES measurements were performed using an MBS A1 spectrometer in St Andrews, which support the conclusions presented here.
  All ARPES data analysis was performed using the {\it peaks} package.~\cite{king2025peaks}

  \

  \noindent{\bf Scanning tunelling microscopy:} In situ STM measurements were performed using a home-built STM which operates in ultra-high vacuum at a temperature of $1.8$ K.\cite{armitage_design_2025}
  Samples are transferred from the MBE to the STM using a vacuum suitcase with typical pressures below $5\times10^{-9}$ mbar before being inserted into the STM.
  The STM tip was made of Pt-Ir wire and prepared on a Au$(111)$ surface by field emission prior to the measurements.
  The bias voltage is applied to the sample with the tip at virtual ground.
  All topographies were taken in constant current mode.
  Differential conductance measurements were performed using a lock-in amplifier with frequency of $371$ Hz, in which the voltage modulation is applied to the sample bias while recording the current.
Details of the QPI data processing are further provided in Supplementary Information, Section V).
Calculations of the quasi-particle interference and unfolded spectral functions have been carried out using calcQPI\cite{wahl_calcqpi_2025}.

\

\noindent{\bf Low-energy electron diffraction:} $\mu$-LEED experiments were conducted at the I06-2 offline laser-PEEM/LEEM (photoemission/low energy electron microscope) instrument at Diamond Light Source, UK.
The room-temperature $\mu$-LEED data was measured from a single HOPG domain as determined from dark-field imaging.
The $\mu$-LEED measurements were performed using a $10\,\mu\text{m}$ illumination aperture, a high voltage of 12\,kV and a start voltage of 22\,V.

\

\noindent{\bf Tight binding model:} the tight binding models used in this work were adapted from the work presented by Kaneko {\it et al.}~\cite{kaneko_exciton-phonon_2018} which follow a Slater-Koster scheme~\cite{slater1954simplified}.
The calculation is outlined in more detail in Supplementary Information, Section IV. In short:

We kept all the nearest-neighbour transfer integral terms unchanged, but modified the on-site energy of Se $4p$ orbitals from $2.171$~eV to $-2.4$~eV to better reproduce the experiment.
We also include the Se $4p$ spin-orbit coupling energy of $\lambda_\mathrm{SOC}=0.105$~eV.\cite{watson_strong-coupling_2020}
The electron-phonon coupling is introduced into the model by taking into account changes to the transfer integral associated with phonon modes, in which we use derivatives of the $d$-$p$ hopping as demonstrated by Motizuki {\it et al.}~\cite{motizuki1986microscopic}
This requires us to introduce the dimensionless term $\alpha$ which determines the strength of the electron-phonon coupling.
The constant can be considered a tunable parameter, and in this work we set $\alpha = 0.3$ which was found to best reproduce experimental data.
Related to the lattice distortion, the magnitude of atomic displacement $u_{\mathbf{q}_i}$ is defined for each phonon mode, providing an order parameter $u=(u_{\mathbf{q}_1},u_{\mathbf{q}_2},u_{\mathbf{q}_3})$ encoding the $\mathbf{q}$-vector structure of the CDW.
In the calculations of Fig.~\ref{fig:flat_bands} and Fig.~\ref{fig:3q_qpi}, we used $u=(u_0,u_0,u_0)$ with $u_0=0.012a$ where $a$ is the normal state lattice constant of TiSe$_{2}$.

To treat the CDW-\moire potential, we consider the enlarged cell with periodicity $8a_{\mathrm{CDW}}\times8a_{\mathrm{CDW}}$, totalling a unit cell with 762 atoms and 5632 orbitals with the inclusion of the spin-orbit coupling.
The potential is introduced via the real space scalar potential:
\begin{equation}
  V(\mathbf{r})=2\sum_{i=1}^{3} V_{i}~\mathrm{cos}(\mathbf{q}_{\mathrm{m},i}^{CDW} \cdot \mathbf{r}),
  \label{eq:moire_potential}
\end{equation}
where $\mathbf{q}_{\mathrm{m},i}^{CDW}$ is the reciprocal $\mathbf{q}$-vector with magnitude of the CDW-\moire potential.
For each $\mathbf{q}_{\mathrm{m},i}^{CDW}$, we define the parameter $(V_1,V_2,V_3)$ which describes the structure of the potential, taking $(V_1,V_2,V_3)=(V_0,V_0,V_0)$ with $V_0=8~\mathrm{meV}$ for $u_0=0.012a$.

\bibliographystyle{naturemag}

\begin{thebibliography}{}
\expandafter\ifx\csname url\endcsname\relax
  \def\url#1{\texttt{#1}}\fi
\expandafter\ifx\csname urlprefix\endcsname\relax\def\urlprefix{URL }\fi
\providecommand{\bibinfo}[2]{#2}
\providecommand{\eprint}[2][]{\url{#2}}

\end{thebibliography}


\begin{thebibliography}{10}
\expandafter\ifx\csname url\endcsname\relax
  \def\url#1{\texttt{#1}}\fi
\expandafter\ifx\csname urlprefix\endcsname\relax\def\urlprefix{URL }\fi
\providecommand{\bibinfo}[2]{#2}
\providecommand{\eprint}[2][]{\url{#2}}

\bibitem{cao_correlated_2018}
\bibinfo{author}{Cao, Y.} \emph{et~al.}
\newblock \bibinfo{title}{Correlated insulator behaviour at half-filling in magic-angle graphene superlattices}.
\newblock \emph{\bibinfo{journal}{Nature}} \textbf{\bibinfo{volume}{556}}, \bibinfo{pages}{80--84} (\bibinfo{year}{2018}).

\bibitem{cao_unconventional_2018}
\bibinfo{author}{Cao, Y.} \emph{et~al.}
\newblock \bibinfo{title}{Unconventional superconductivity in magic-angle graphene superlattices}.
\newblock \emph{\bibinfo{journal}{Nature}} \textbf{\bibinfo{volume}{556}}, \bibinfo{pages}{43--50} (\bibinfo{year}{2018}).

\bibitem{andrei_marvels_2021}
\bibinfo{author}{Andrei, E.~Y.} \emph{et~al.}
\newblock \bibinfo{title}{The marvels of moiré materials}.
\newblock \emph{\bibinfo{journal}{Nat. Rev. Mater.}} \textbf{\bibinfo{volume}{6}}, \bibinfo{pages}{201--206} (\bibinfo{year}{2021}).

\bibitem{kennes_moire_2021}
\bibinfo{author}{Kennes, D.~M.} \emph{et~al.}
\newblock \bibinfo{title}{Moiré heterostructures as a condensed-matter quantum simulator}.
\newblock \emph{\bibinfo{journal}{Nat. Phys,}} \textbf{\bibinfo{volume}{17}}, \bibinfo{pages}{155--163} (\bibinfo{year}{2021}).

\bibitem{mak_semiconductor_2022}
\bibinfo{author}{Mak, K.~F.} \& \bibinfo{author}{Shan, J.}
\newblock \bibinfo{title}{Semiconductor moiré materials}.
\newblock \emph{\bibinfo{journal}{Nat. Nanotechnol.}} \textbf{\bibinfo{volume}{17}}, \bibinfo{pages}{686--695} (\bibinfo{year}{2022}).

\bibitem{nuckolls_microscopic_2024}
\bibinfo{author}{Nuckolls, K.~P.} \& \bibinfo{author}{Yazdani, A.}
\newblock \bibinfo{title}{A microscopic perspective on moiré materials}.
\newblock \emph{\bibinfo{journal}{Nat. Rev. Mat.}} \textbf{\bibinfo{volume}{9}}, \bibinfo{pages}{460--480} (\bibinfo{year}{2024}).

\bibitem{geim_van_2013}
\bibinfo{author}{Geim, A.~K.} \& \bibinfo{author}{Grigorieva, I.~V.}
\newblock \bibinfo{title}{Van der {Waals} heterostructures}.
\newblock \emph{\bibinfo{journal}{Nature}} \textbf{\bibinfo{volume}{499}}, \bibinfo{pages}{419--425} (\bibinfo{year}{2013}).

\bibitem{bistritzer_moire_2011}
\bibinfo{author}{Bistritzer, R.} \& \bibinfo{author}{MacDonald, A.~H.}
\newblock \bibinfo{title}{Moiré bands in twisted double-layer graphene}.
\newblock \emph{\bibinfo{journal}{Proc. Natl. Acad. Sci. U.S.A.}} \textbf{\bibinfo{volume}{108}}, \bibinfo{pages}{12233--12237} (\bibinfo{year}{2011}).

\bibitem{wu_topological_2017}
\bibinfo{author}{Wu, F.}, \bibinfo{author}{Lovorn, T.} \& \bibinfo{author}{MacDonald, A.}
\newblock \bibinfo{title}{Topological {Exciton} {Bands} in {Moir}{\'e} {Heterojunctions}}.
\newblock \emph{\bibinfo{journal}{Phys. Rev. Lett.}} \textbf{\bibinfo{volume}{118}}, \bibinfo{pages}{147401} (\bibinfo{year}{2017}).

\bibitem{utama_visualization_2021}
\bibinfo{author}{Utama, M. I.~B.} \emph{et~al.}
\newblock \bibinfo{title}{Visualization of the flat electronic band in twisted bilayer graphene near the magic angle twist}.
\newblock \emph{\bibinfo{journal}{Nat. Phys.}} \textbf{\bibinfo{volume}{17}}, \bibinfo{pages}{184--188} (\bibinfo{year}{2021}).

\bibitem{lisi_observation_2021}
\bibinfo{author}{Lisi, S.} \emph{et~al.}
\newblock \bibinfo{title}{Observation of flat bands in twisted bilayer graphene}.
\newblock \emph{\bibinfo{journal}{Nat. Phys.}} \textbf{\bibinfo{volume}{17}}, \bibinfo{pages}{189--193} (\bibinfo{year}{2021}).

\bibitem{angeli__2021}
\bibinfo{author}{Angeli, M.} \& \bibinfo{author}{MacDonald, A.~H.}
\newblock \bibinfo{title}{{$\Gamma$} valley transition metal dichalcogenide moiré bands}.
\newblock \emph{\bibinfo{journal}{Proc. Natl. Acad. Sci. U.S.A.}} \textbf{\bibinfo{volume}{118}}, \bibinfo{pages}{e2021826118} (\bibinfo{year}{2021}).

\bibitem{pei_observation_2022}
\bibinfo{author}{Pei, D.} \emph{et~al.}
\newblock \bibinfo{title}{Observation of $\gamma$-valley moir{\'e} bands and emergent hexagonal lattice in twisted transition metal dichalcogenides}.
\newblock \emph{\bibinfo{journal}{Phys. Rev. X}} \textbf{\bibinfo{volume}{12}}, \bibinfo{pages}{021065} (\bibinfo{year}{2022}).

\bibitem{gatti_flat_2023}
\bibinfo{author}{Gatti, G.} \emph{et~al.}
\newblock \bibinfo{title}{Flat $\gamma$ moir{\'e} bands in twisted bilayer {WSe}$_{2}$}.
\newblock \emph{\bibinfo{journal}{Phys. Rev. Lett.}} \textbf{\bibinfo{volume}{131}}, \bibinfo{pages}{046401} (\bibinfo{year}{2023}).

\bibitem{chen_evidence_2019}
\bibinfo{author}{Chen, G.} \emph{et~al.}
\newblock \bibinfo{title}{Evidence of a gate-tunable {Mott} insulator in a trilayer graphene moiré superlattice}.
\newblock \emph{\bibinfo{journal}{Nat. Phys.}} \textbf{\bibinfo{volume}{15}}, \bibinfo{pages}{237--241} (\bibinfo{year}{2019}).

\bibitem{regan_mott_2020}
\bibinfo{author}{Regan, E.~C.} \emph{et~al.}
\newblock \bibinfo{title}{Mott and generalized {Wigner} crystal states in {WSe}$_{2}$/{WS}$_{2}$ moir{\'e} superlattices}.
\newblock \emph{\bibinfo{journal}{Nature}} \textbf{\bibinfo{volume}{579}}, \bibinfo{pages}{359--363} (\bibinfo{year}{2020}).

\bibitem{tang_simulation_2020}
\bibinfo{author}{Tang, Y.} \emph{et~al.}
\newblock \bibinfo{title}{Simulation of {Hubbard} model physics in {WSe}$_{2}$/{WS}$_{2}$ moiré superlattices}.
\newblock \emph{\bibinfo{journal}{Nature}} \textbf{\bibinfo{volume}{579}}, \bibinfo{pages}{353--358} (\bibinfo{year}{2020}).

\bibitem{li_quantum_2021}
\bibinfo{author}{Li, T.} \emph{et~al.}
\newblock \bibinfo{title}{Quantum anomalous {Hall} effect from intertwined moiré bands}.
\newblock \emph{\bibinfo{journal}{Nature}} \textbf{\bibinfo{volume}{600}}, \bibinfo{pages}{641--646} (\bibinfo{year}{2021}).

\bibitem{sharpe_emergent_2019}
\bibinfo{author}{Sharpe, A.~L.} \emph{et~al.}
\newblock \bibinfo{title}{Emergent ferromagnetism near three-quarters filling in twisted bilayer graphene}.
\newblock \emph{\bibinfo{journal}{Science}} \textbf{\bibinfo{volume}{365}}, \bibinfo{pages}{605--608} (\bibinfo{year}{2019}).

\bibitem{chen_tunable_2020}
\bibinfo{author}{Chen, G.} \emph{et~al.}
\newblock \bibinfo{title}{Tunable correlated {Chern} insulator and ferromagnetism in a moiré superlattice}.
\newblock \emph{\bibinfo{journal}{Nature}} \textbf{\bibinfo{volume}{579}}, \bibinfo{pages}{56--61} (\bibinfo{year}{2020}).

\bibitem{serlin_intrinsic_2020}
\bibinfo{author}{Serlin, M.} \emph{et~al.}
\newblock \bibinfo{title}{Intrinsic quantized anomalous {Hall} effect in a moiré heterostructure}.
\newblock \emph{\bibinfo{journal}{Science}} \textbf{\bibinfo{volume}{367}}, \bibinfo{pages}{900--903} (\bibinfo{year}{2020}).

\bibitem{tschirhart_imaging_2021}
\bibinfo{author}{Tschirhart, C.~L.} \emph{et~al.}
\newblock \bibinfo{title}{Imaging orbital ferromagnetism in a moiré {Chern} insulator}.
\newblock \emph{\bibinfo{journal}{Science}} \textbf{\bibinfo{volume}{372}}, \bibinfo{pages}{1323--1327} (\bibinfo{year}{2021}).

\bibitem{yankowitz_tuning_2019}
\bibinfo{author}{Yankowitz, M.} \emph{et~al.}
\newblock \bibinfo{title}{Tuning superconductivity in twisted bilayer graphene}.
\newblock \emph{\bibinfo{journal}{Science}} \textbf{\bibinfo{volume}{363}}, \bibinfo{pages}{1059--1064} (\bibinfo{year}{2019}).

\bibitem{lu_superconductors_2019}
\bibinfo{author}{Lu, X.} \emph{et~al.}
\newblock \bibinfo{title}{Superconductors, orbital magnets and correlated states in magic-angle bilayer graphene}.
\newblock \emph{\bibinfo{journal}{Nature}} \textbf{\bibinfo{volume}{574}}, \bibinfo{pages}{653--657} (\bibinfo{year}{2019}).

\bibitem{saito_hofstadter_2021}
\bibinfo{author}{Saito, Y.} \emph{et~al.}
\newblock \bibinfo{title}{Hofstadter subband ferromagnetism and symmetry-broken {Chern} insulators in twisted bilayer graphene}.
\newblock \emph{\bibinfo{journal}{Nat. Phys.}} \textbf{\bibinfo{volume}{17}}, \bibinfo{pages}{478--481} (\bibinfo{year}{2021}).

\bibitem{liu_visualizing_2022}
\bibinfo{author}{Liu, X.} \emph{et~al.}
\newblock \bibinfo{title}{Visualizing broken symmetry and topological defects in a quantum {Hall} ferromagnet}.
\newblock \emph{\bibinfo{journal}{Science}} \textbf{\bibinfo{volume}{375}}, \bibinfo{pages}{321--326} (\bibinfo{year}{2022}).

\bibitem{yu_correlated_2022}
\bibinfo{author}{Yu, J.} \emph{et~al.}
\newblock \bibinfo{title}{Correlated {Hofstadter} spectrum and flavour phase diagram in magic-angle twisted bilayer graphene}.
\newblock \emph{\bibinfo{journal}{Nat. Phys.}} \textbf{\bibinfo{volume}{18}}, \bibinfo{pages}{825--831} (\bibinfo{year}{2022}).

\bibitem{di_salvo_electronic_1976}
\bibinfo{author}{Di~Salvo, F.~J.}, \bibinfo{author}{Moncton, D.~E.} \& \bibinfo{author}{Waszczak, J.~V.}
\newblock \bibinfo{title}{Electronic properties and superlattice formation in the semimetal {TiSe}$_{2}$}.
\newblock \emph{\bibinfo{journal}{Phys. Rev. B}} \textbf{\bibinfo{volume}{14}}, \bibinfo{pages}{4321--4328} (\bibinfo{year}{1976}).

\bibitem{peng_molecular_2015}
\bibinfo{author}{Peng, J.-P.} \emph{et~al.}
\newblock \bibinfo{title}{Molecular beam epitaxy growth and scanning tunneling microscopy study of {TiSe}$_{2}$ ultrathin films}.
\newblock \emph{\bibinfo{journal}{Phys. Rev. B}} \textbf{\bibinfo{volume}{91}}, \bibinfo{pages}{121113} (\bibinfo{year}{2015}).

\bibitem{chen_charge_2015}
\bibinfo{author}{Chen, P.} \emph{et~al.}
\newblock \bibinfo{title}{Charge density wave transition in single-layer titanium diselenide}.
\newblock \emph{\bibinfo{journal}{Nat. Commun.}} \textbf{\bibinfo{volume}{6}}, \bibinfo{pages}{8943} (\bibinfo{year}{2015}).

\bibitem{watson_strong-coupling_2020}
\bibinfo{author}{Watson, M.~D.} \emph{et~al.}
\newblock \bibinfo{title}{Strong-coupling charge density wave in monolayer {TiSe}$_{2}$}.
\newblock \emph{\bibinfo{journal}{2D Materials}} \textbf{\bibinfo{volume}{8}}, \bibinfo{pages}{015004} (\bibinfo{year}{2020}).

\bibitem{buchberger_persistence_2025}
\bibinfo{author}{Buchberger, S.} \emph{et~al.}
\newblock \bibinfo{title}{Persistence of charge ordering instability to coulomb engineering in the excitonic insulator candidate {TiSe$_2$}}.
\newblock \emph{\bibinfo{journal}{Phys. Rev. X}} \textbf{\bibinfo{volume}{15}}, \bibinfo{pages}{041028} (\bibinfo{year}{2025}).

\bibitem{zeller_what_2014}
\bibinfo{author}{Zeller, P.} \& \bibinfo{author}{Günther, S.}
\newblock \bibinfo{title}{What are the possible moiré patterns of graphene on hexagonally packed surfaces? {Universal} solution for hexagonal coincidence lattices, derived by a geometric construction}.
\newblock \emph{\bibinfo{journal}{New J. Phys.}} \textbf{\bibinfo{volume}{16}}, \bibinfo{pages}{083028} (\bibinfo{year}{2014}).

\bibitem{mo2026resonant}
\bibinfo{author}{Mo, S.} \emph{et~al.}
\newblock \bibinfo{title}{Resonant interlayer coupling in {NbSe$_{2}$}-graphite epitaxial moir{\'e} superlattices}.
\newblock \emph{\bibinfo{journal}{Adv. Mater.}} \textbf{\bibinfo{volume}{38}}, \bibinfo{pages}{e11262} (\bibinfo{year}{2026}).

\bibitem{rossnagel_charge-density-wave_2002}
\bibinfo{author}{Rossnagel, K.}, \bibinfo{author}{Kipp, L.} \& \bibinfo{author}{Skibowski, M.}
\newblock \bibinfo{title}{Charge-density-wave phase transition in 1{T}-{TiSe$_2$}: {Excitonic} insulator versus band-type {Jahn}-{Teller} mechanism}.
\newblock \emph{\bibinfo{journal}{Phys. Rev. B}} \textbf{\bibinfo{volume}{65}}, \bibinfo{pages}{235101} (\bibinfo{year}{2002}).

\bibitem{cercellier_evidence_2007}
\bibinfo{author}{Cercellier, H.} \emph{et~al.}
\newblock \bibinfo{title}{Evidence for an excitonic insulator phase in {1T} {TiSe}$_{2}$}.
\newblock \emph{\bibinfo{journal}{Phys. Rev. Lett.}} \textbf{\bibinfo{volume}{99}}, \bibinfo{pages}{146403} (\bibinfo{year}{2007}).

\bibitem{watson_orbital-and_2019}
\bibinfo{author}{Watson, M.~D.} \emph{et~al.}
\newblock \bibinfo{title}{Orbital- and k$_{z}$-selective hybridization of se $4p$ and ti $3d$ states in the charge density wave phase of {TiSe}$_{2}$}.
\newblock \emph{\bibinfo{journal}{Phys. Rev. Lett.}} \textbf{\bibinfo{volume}{122}} (\bibinfo{year}{2019}).

\bibitem{kaneko_exciton-phonon_2018}
\bibinfo{author}{Kaneko, T.}, \bibinfo{author}{Ohta, Y.} \& \bibinfo{author}{Yunoki, S.}
\newblock \bibinfo{title}{Exciton-phonon cooperative mechanism of the triple-$q$ charge-density-wave and antiferroelectric electron polarization in {TiSe}$_{2}$}.
\newblock \emph{\bibinfo{journal}{Phys. Rev. B}} \textbf{\bibinfo{volume}{97}}, \bibinfo{pages}{155131} (\bibinfo{year}{2018}).

\bibitem{bianco_electronic_2015}
\bibinfo{author}{Bianco, R.}, \bibinfo{author}{Calandra, M.} \& \bibinfo{author}{Mauri, F.}
\newblock \bibinfo{title}{Electronic and vibrational properties of {TiSe}$_{2}$ in the charge-density-wave phase from first principles}.
\newblock \emph{\bibinfo{journal}{Phys. Rev. B}} \textbf{\bibinfo{volume}{92}}, \bibinfo{pages}{094107} (\bibinfo{year}{2015}).

\bibitem{monney_spontaneous_2009}
\bibinfo{author}{Monney, C.} \emph{et~al.}
\newblock \bibinfo{title}{Spontaneous exciton condensation in {1T}-{TiSe}$_{2}$: {BCS}-like approach}.
\newblock \emph{\bibinfo{journal}{Phys. Rev. B}} \textbf{\bibinfo{volume}{79}}, \bibinfo{pages}{045116} (\bibinfo{year}{2009}).

\bibitem{antonelli_orbital-selective_2022}
\bibinfo{author}{Antonelli, T.} \emph{et~al.}
\newblock \bibinfo{title}{Orbital-selective band hybridisation at the charge density wave transition in monolayer {TiTe}$_{2}$}.
\newblock \emph{\bibinfo{journal}{npj Quantum Mater.}} \textbf{\bibinfo{volume}{7}} (\bibinfo{year}{2022}).

\bibitem{claassen_ultra-strong_2022}
\bibinfo{author}{Claassen, M.}, \bibinfo{author}{Xian, L.}, \bibinfo{author}{Kennes, D.~M.} \& \bibinfo{author}{Rubio, A.}
\newblock \bibinfo{title}{Ultra-strong spin–orbit coupling and topological moiré engineering in twisted {ZrS}$_{2}$ bilayers}.
\newblock \emph{\bibinfo{journal}{Nat. Commun.}} \textbf{\bibinfo{volume}{13}}, \bibinfo{pages}{4915} (\bibinfo{year}{2022}).

\bibitem{joe2014emergence}
\bibinfo{author}{Joe, Y.~I.} \emph{et~al.}
\newblock \bibinfo{title}{Emergence of charge density wave domain walls above the superconducting dome in {1T}-{TiSe}$_{2}$}.
\newblock \emph{\bibinfo{journal}{Nat. Phys}} \textbf{\bibinfo{volume}{10}}, \bibinfo{pages}{421--425} (\bibinfo{year}{2014}).

\bibitem{kogar2017observation}
\bibinfo{author}{Kogar, A.} \emph{et~al.}
\newblock \bibinfo{title}{Observation of a charge density wave incommensuration near the superconducting dome in {Cu}$_{x}${TiSe}$_{2}$}.
\newblock \emph{\bibinfo{journal}{Phys. Rev. Lett.}} \textbf{\bibinfo{volume}{118}}, \bibinfo{pages}{027002} (\bibinfo{year}{2017}).

\bibitem{yan2017influence}
\bibinfo{author}{Yan, S.} \emph{et~al.}
\newblock \bibinfo{title}{Influence of domain walls in the incommensurate charge density wave state of {Cu} intercalated {1T}-{TiSe}$_{2}$}.
\newblock \emph{\bibinfo{journal}{Phys. Rev. Lett.}} \textbf{\bibinfo{volume}{118}}, \bibinfo{pages}{106405} (\bibinfo{year}{2017}).

\bibitem{munoz2025doping}
\bibinfo{author}{Mu{\~n}oz-Segovia, D.}, \bibinfo{author}{Venderbos, J.~W.}, \bibinfo{author}{Grushin, A.~G.} \& \bibinfo{author}{de~Juan, F.}
\newblock \bibinfo{title}{Doping-induced nematic and stripe orders within the charge density wave state of {TiSe}$_{2}$}.
\newblock \emph{\bibinfo{journal}{Phys. Rev. B}} \textbf{\bibinfo{volume}{112}}, \bibinfo{pages}{165119} (\bibinfo{year}{2025}).

\bibitem{wan_near-commensuration_2026}
\bibinfo{author}{Wan, W.} \emph{et~al.}
\newblock \bibinfo{title}{Near-commensuration from {Intertwined} {Charge} {Density} {Waves} in {Single}-{Layer} {TiSe}$_{\textrm{2}}$}.
\newblock \emph{\bibinfo{journal}{Nano Lett.}} \textbf{\bibinfo{volume}{26}}, \bibinfo{pages}{816--822} (\bibinfo{year}{2026}).

\bibitem{li_controlling_2015}
\bibinfo{author}{Li, L.~J.} \emph{et~al.}
\newblock \bibinfo{title}{Controlling many-body states by the electric-field effect in a two-dimensional material}.
\newblock \emph{\bibinfo{journal}{Nature}} \textbf{\bibinfo{volume}{529}}, \bibinfo{pages}{185--189} (\bibinfo{year}{2015}).

\bibitem{morosan_superconductivity_2006}
\bibinfo{author}{Morosan, E.} \emph{et~al.}
\newblock \bibinfo{title}{Superconductivity in {Cu}$_{x}${TiSe}$_{2}$}.
\newblock \emph{\bibinfo{journal}{Nat. Phys.}} \textbf{\bibinfo{volume}{2}}, \bibinfo{pages}{544--550} (\bibinfo{year}{2006}).

\bibitem{bostwick_quasiparticle_2007}
\bibinfo{author}{Bostwick, A.}, \bibinfo{author}{Ohta, T.}, \bibinfo{author}{Seyller, T.}, \bibinfo{author}{Horn, K.} \& \bibinfo{author}{Rotenberg, E.}
\newblock \bibinfo{title}{Quasiparticle dynamics in graphene}.
\newblock \emph{\bibinfo{journal}{Nat. Phys.}} \textbf{\bibinfo{volume}{3}}, \bibinfo{pages}{36--40} (\bibinfo{year}{2007}).

\bibitem{kim_observation_2015}
\bibinfo{author}{Kim, J.} \emph{et~al.}
\newblock \bibinfo{title}{Observation of tunable band gap and anisotropic {Dirac} semimetal state in black phosphorus}.
\newblock \emph{\bibinfo{journal}{Science}} \textbf{\bibinfo{volume}{349}}, \bibinfo{pages}{723--726} (\bibinfo{year}{2015}).

\bibitem{riley_negative_2015}
\bibinfo{author}{Riley, J.~M.} \emph{et~al.}
\newblock \bibinfo{title}{Negative electronic compressibility and tunable spin splitting in {WSe}$_{2}$}.
\newblock \emph{\bibinfo{journal}{Nat. Nanotechnol.}} \textbf{\bibinfo{volume}{10}}, \bibinfo{pages}{1043--+} (\bibinfo{year}{2015}).

\bibitem{antonelli_controlling_2023}
\bibinfo{author}{Antonelli, T.} \emph{et~al.}
\newblock \bibinfo{title}{Controlling the charge density wave transition in single-layer {TiTe}$_{2x}${Se}$_{2(1–x)}$ alloys by band gap engineering}.
\newblock \emph{\bibinfo{journal}{Nano Lett.}} \textbf{\bibinfo{volume}{24}}, \bibinfo{pages}{215--221} (\bibinfo{year}{2023}).

\bibitem{Rajan2024}
\bibinfo{author}{Rajan, A.} \emph{et~al.}
\newblock \bibinfo{title}{Epitaxial growth of large-area monolayers and van der {W}aals heterostructures of transition-metal chalcogenides via assisted nucleation}.
\newblock \emph{\bibinfo{journal}{Adv. Mater.}} \textbf{\bibinfo{volume}{36}}, \bibinfo{pages}{2402254} (\bibinfo{year}{2024}).

\bibitem{rajan_substrate_2025}
\bibinfo{author}{Rajan, A.} \emph{et~al.}
\newblock \bibinfo{title}{Substrate pre-sputtering for layer-by-layer van der waals epitaxy of {2D} materials}.
\newblock \emph{\bibinfo{journal}{APL Materials}} \textbf{\bibinfo{volume}{13}} (\bibinfo{year}{2025}).

\bibitem{king2025peaks}
\bibinfo{author}{King, P. D.~C.} \emph{et~al.}
\newblock \bibinfo{title}{peaks: a python package for analysis of angle-resolved photoemission and related spectroscopies}  (\bibinfo{year}{2025}).
\newblock \bibinfo{note}{ArXiv: 2508.04803 [cond-mat]}, \eprint{2508.04803}.

\bibitem{armitage_design_2025}
\bibinfo{author}{Armitage, O.} \emph{et~al.}
\newblock \bibinfo{title}{Design for a 1 {K} pot for a low-temperature ultra-high vacuum scanning tunneling microscope}.
\newblock \emph{\bibinfo{journal}{Rev. Sci. Instrum.}} \textbf{\bibinfo{volume}{96}}, \bibinfo{pages}{123906} (\bibinfo{year}{2025}).

\bibitem{wahl_calcqpi_2025}
\bibinfo{author}{Wahl, P.}, \bibinfo{author}{Rhodes, L.~C.} \& \bibinfo{author}{Marques, C.~A.}
\newblock \bibinfo{title}{{calcQPI}: {A} versatile tool to simulate quasiparticle interference}.
\newblock \emph{\bibinfo{journal}{SciPost Phys. Codebases}} \bibinfo{pages}{61} (\bibinfo{year}{2025}).

\bibitem{slater1954simplified}
\bibinfo{author}{Slater, J.~C.} \& \bibinfo{author}{Koster, G.~F.}
\newblock \bibinfo{title}{Simplified lcao method for the periodic potential problem}.
\newblock \emph{\bibinfo{journal}{Phys. Rev.}} \textbf{\bibinfo{volume}{94}}, \bibinfo{pages}{1498} (\bibinfo{year}{1954}).

\bibitem{motizuki1986microscopic}
\bibinfo{author}{Motizuki, K.} \& \bibinfo{author}{Suzuki, N.}
\newblock \bibinfo{title}{Microscopic theory of structural phase transitions in layered transitional-metal compounds}.
\newblock In \emph{\bibinfo{booktitle}{Structural Phase Transitions in Layered Transition Metal Compounds}}, \bibinfo{pages}{1--133} (\bibinfo{publisher}{Springer}, \bibinfo{year}{1986}).

\end{thebibliography}

\foreach \x in {1,2,3,4,5,6,7,8}
{%
\begin{figure*}[!ht]
  \centering
  \includegraphics[
  width=\textwidth,
  page=\x,
  trim=1.5cm 1cm 1.5cm 1.5cm,
  clip
]{supp.pdf}
\end{figure*}
}

\end{document}